\documentclass[twocolumn,showpacs,superscriptaddress,11pt]{revtex4-1} 
\usepackage{amsmath,amssymb,graphicx}
\begin{document}

\title{A sub nrad beam pointing monitoring and stabilization system for controlling input beam jitter in GW interferometers. }

\author{B. Canuel}\email{presently at Institut d'Optique d'Aquitaine LP2N - Laboratoire Photonique, Num\'erique et Nanosciences 2 All\'ee Ren\'e Laroumagne 33405 TALENCE CEDEX}
\affiliation{European Gravitational Observatory (EGO), I-56021 Cascina (Pi), Italy}
\author{E. Genin}
\affiliation{European Gravitational Observatory (EGO), I-56021 Cascina (Pi), Italy}
\author{M. Mantovani}\email{Corresponding author: maddalena.mantovani@ego-gw.it}
\affiliation{European Gravitational Observatory (EGO), I-56021 Cascina (Pi), Italy}
\author{J. Marque}
\affiliation{European Gravitational Observatory (EGO), I-56021 Cascina (Pi), Italy}
\author{P. Ruggi}
\affiliation{European Gravitational Observatory (EGO), I-56021 Cascina (Pi), Italy}
\author{M. Tacca} \email{presently at APC, AstroParticule et Cosmologie, Universit\'e Paris Diderot, F-75205 Paris, France}
\affiliation{European Gravitational Observatory (EGO), I-56021 Cascina (Pi), Italy}

\begin{abstract}
                       In this paper a simple and very effective control system to monitor and suppress the beam jitter noise at the input of an optical system, called  \textit{Beam Pointing Control} (BPC) system, will be described showing the theoretical principle and an experimental demonstration for the application of large scale gravitational wave interferometers, in particular for the Advanced Virgo detector.\\
For this purpose the requirements for the control accuracy and the sensing noise will be computed by taking into account the Advanced Virgo optical configuration and the outcomes will be compared with the experimental measurement obtained in the laboratory.
The system has shown unprecedented performance in terms of control accuracy and sensing noise. The BPC system has achieved a control accuracy  of $\sim 10^{-8}$~rad for the tilt and $\sim 10^{-7}$~m for the shift and a sensing noise of less than 1~$n$rad$/\sqrt{Hz}$ resulting compliant with the Advance Virgo gravitational wave interferometer requirements.
\end{abstract}
\maketitle
\section{Introduction}
The beam pointing noise is an issue impacting the performance of various optical systems. Complexity of high power lasers systems usually creates rather unstable pointing performance which can be a major issue for their applications \cite{TW, Femto, KrF}. Pointing stabilization can be necessary in many areas of physics such as atom optical trapping \cite{opticaltrapping}, microscopy \cite{stabconf} or free-space laser communication \cite{lasercomm}.
It can also be a major source of technical noise in km-scale interferometric gravitational wave (GW) detectors. Indeed, it was identified at an early stage that in case of geometrical asymmetries between the arms of the interferometer (ITF), created by spurious misalignments of ITF optics, the input beam jitter in the detector frequency bandwidth (10Hz-10kHz) creates a phase noise directly affecting detector sensitivity~\cite{misalgnVirgo, misalgnLigo}. In order to mitigate this effect, the input beam jitter is filtered out by means of a mode cleaning cavity. Despite this precaution, the commissioning of the first generation of GW detectors showed that the input beam fluctuations can be a limiting technical noise especially for frequencies around 100~Hz. This could be even worse for the second generation of detectors \cite{TDR, ALIGO} where jitter specifications at low frequency become even more stringent because of the radiation pressure effects~\cite{TDR}. In addition, the fluctuations of beam pointing at low frequencies (DC-10Hz) can impact the ITF lock accuracy, which indirectly degrades the detector sensitivity \cite{vsr1}. In this paper the specifications of input beam jitter over the whole frequency band of the $2^{nd}$ generation detector Advanced Virgo (DC-10kHz) will be evaluated. A system to monitor the input beam jitter and to estimate the compliance of measured jitter noise with respect to the specifications will then be described.
Finally the control system, the \textit{Beam Pointing Control} (BPC) system, used to deal with the low frequency large fluctuation of the pointing, due to air flux and thermal drift will be detailed, showing also the control performance.

\section{Beam pointing requirements for the input beam of GW interferometer experiments}\label{theory}
\begin{figure}[htp]
    \centering
    \includegraphics[width=0.5\textwidth,clip]{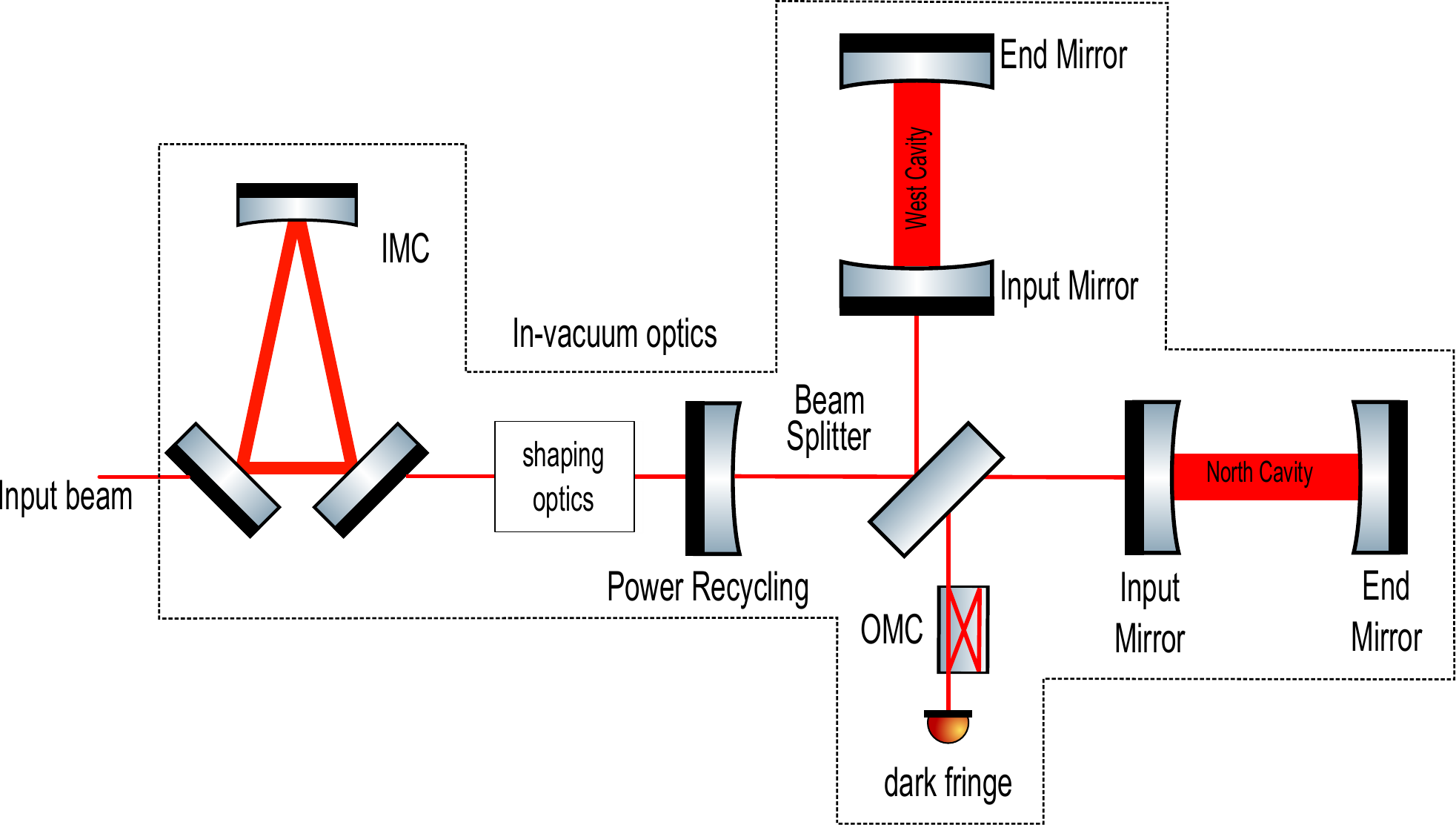}
    \caption{Scheme of advanced laser interferometer GW detectors.}
    \label{OptLay}
\end{figure}

Figure~\ref{OptLay} shows the general scheme of the Advanced Virgo double recycled Michelson interferometer. 
The laser source and some part of the input optics system are placed on a set of in-air optical benches where they experience jitter noise due to spurious electro-optical effects, vibrations, thermal fluctuations and air flux variations before being sent through the suspended in-vacuum optical system. After entering the vacuum system, the beam passes through a triangular mode cleaning cavity (IMC) and is then mode matched onto the long arm Fabry-Perot cavities using a set of shaping optics. The gravitational wave signal is obtained from the dark fringe of the Michelson which is filtered by means of an output mode cleaner cavity (OMC).\\
The beam jitter at the input of the IMC impacts the detector performance in different ways depending on the perturbation frequency. In this section, the calculation of the jitter requirements in the detection frequency band (from 10 Hz to kHz) and at lower frequencies (DC-10Hz) will be carried out. 

\subsection{Calculation of beam jitter requirements for Advanced Virgo in the detection frequency band.}
\label{sec:highfrequreq}

In this section the method that has been used to compute the beam jitter requirements for Advanced Virgo at the interferometer (ITF) input, i.e. at the level of the Power Recycling Mirror (PR), will be described. The requirements at the IMC input will then be determined taking into account the IMC and the shaping optics properties.\\

The jitter of the ITF input beam couples into the dark fringe signal through optical asymmetries between the two arms of the interferometer and therefore it mainly couples through the residual RMS tilt motion of the core optics~\cite{misalgnLigo}. \\

\begin{table}[htp]
\begin{center}
\begin{tabular}{|l|c|}
\hline
Degree of freedom & Requirement (nrad RMS)\\
\hline
(+)-mode & 2 \\
\hline
(-)-mode & 110 \\
\hline
PRM & 25 \\
\hline
SRM & 280 \\
\hline
BS & 35 \\
\hline
\end{tabular}
\end{center}
\caption{\label{tab_alignmentRMS} Angular requirements for Advanced Virgo for the arm cavity modes (the (+) and (-) modes) and for the central interferometer mirrors, Beam Splitter (BS), Power Recycling (PRM) and Signal Recycling (SRM) mirrors.}
\end{table}

The jitter noise requirement can be obtained by evaluating the power Transfer Function $TF_{TEM_{01}}$ between an  Hermite-Gauss Transverse Electro-Magnetic field of order 1 at the input of the interferometer and the fundamental Gaussian beam at the output of the interferometer of the carrier field and then converting this power noise into strain sensitivity by modeling the response of the interferometer $TF_{h/W}$ from Watt to $h$. 

In this computation it has been assumed that the main channel for the jitter coupling is the carrier field. This is due to the fact that the carrier field at the dark fringe is much more powerful with respect to the sideband fields since they are drastically filtered by the Output Mode Cleaner cavity. Moreover, the reason why only the fundamental mode at the output of the interferometer can be considered in the computation of the jitter noise is because the Higher Order Modes will be filtered by the Output Mode Cleaner, clearly assuming a perfect alignment of the Output Mode Cleaner with respect to the dark fringe field.\\

The Transfer Function $TF_{TEM_{01}}$ can be computed analytically, as has been done in~\cite{Mueller05}\cite{T0900142-v2} where the jitter requirements for Advanced LIGO are derived, but it gets rather complicated when all asymmetries have to be taken into account. For this reason, a numerical calculation has been carried out using the frequency domain simulation tool \textit{Finesse}~\cite{Freise04} in order to compute directly the transfer function $TF_{TEM_{01}}$ for any kind of optical configurations and interferometer defects.\\

The reference interferometer configuration~\cite{TDR}, for which the operating point has been optimized for Binary Neutron Star detection, has then been modeled by adding the static misalignments foreseen for the Advanced Virgo core optics, shown in Table \ref{tab_alignmentRMS}~\cite{AAdes}, by choosing the combination which yields the worst case scenario, i.e. the one which maximizes the power transfer function.\\
The effect of the jitter noise will then be converted into sensitivity by evaluating the response of the interferometer $TF_{W/h}$, from $h$ to Watt,  which has been computed with the \textit{Optickle} simulation software \cite{Optickle_homepage} to take into account the radiation pressure effect. \\
The requirement for the input beam jitter is then computed as:
\begin{equation}\label{eq:requirements}
S_{Jitter} = \frac{h_{AdV}}{10} \times \frac{TF_{W/ h}}{TF_{TEM_{01}}}
\end{equation}
where $h_{AdV}/10$ is the target Advanced Virgo strain sensitivity taking into account a factor 10 safety margin. The $TF_{TEM_{01}}$ is the outcome of the \textit{Finesse} simulation and $TF_{W/h}$ is the response of the interferometer obtained with \textit{Optickle}.\\
\begin{figure}[htbp]
\begin{center}
\includegraphics[width=0.5\textwidth,clip]{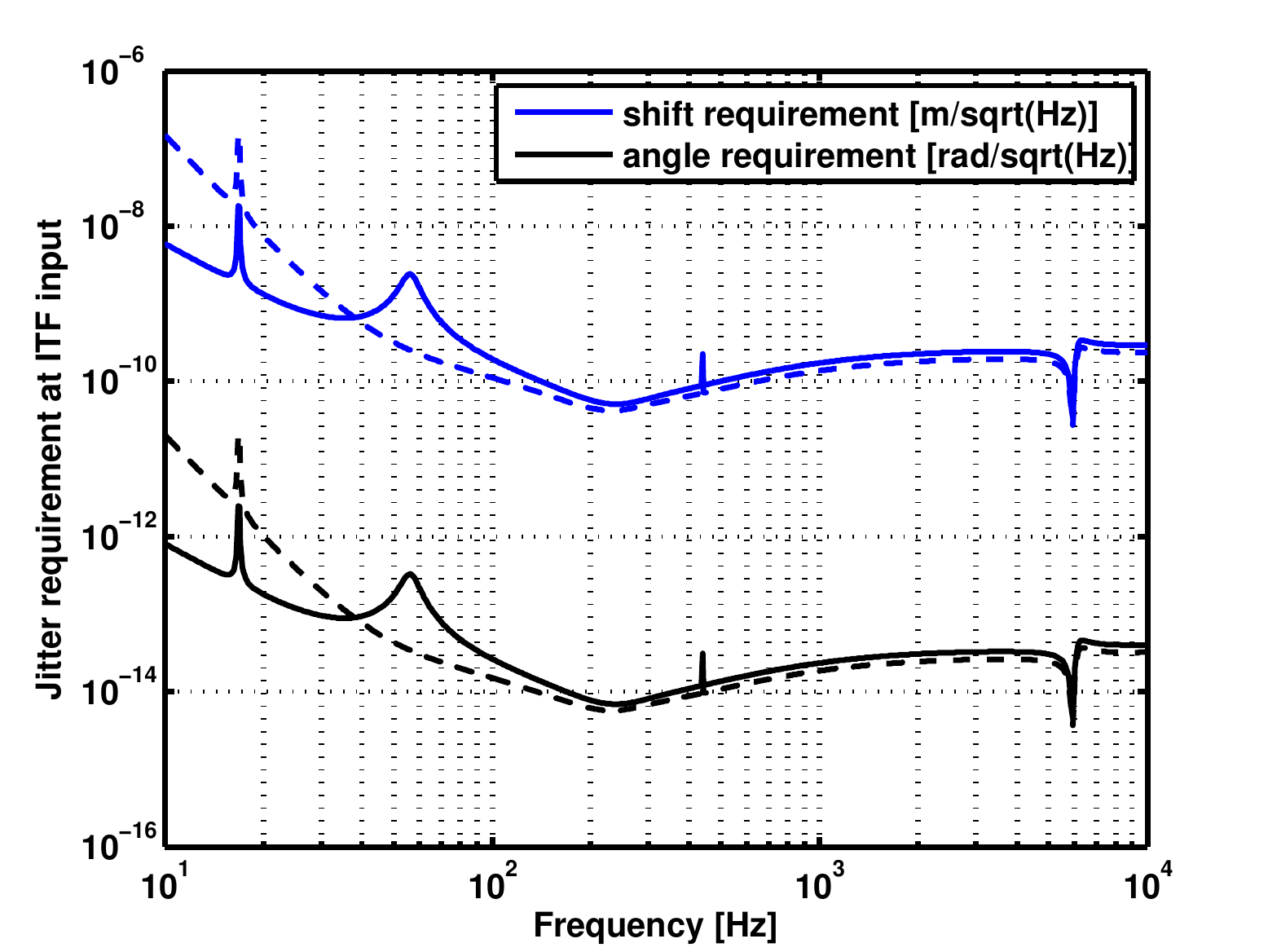}
\caption{\label{fig_jitter_requirements} Advanced Virgo Jitter requirements at the input of the interferometer, solid curves, for the tilt and shift d.o.f.. The dashed curves represent the jitter requirements in case the radiation pressure effects have not taken into account in the modeling showing requirements much more relaxed at low frequencies, of about a factor $\sim$30 at 10~Hz. }
\label{default}
\end{center}
\end{figure}
It is worth noting that two different packages had to be used for this computation since \textit{Finesse} does not take into account the radiation pressure effect which plays a key role in the second generation interferometers due to the high circulating power, making the requirement more stringent at low frequency. On the other hand \textit{Optickle} does not allow to introduce asymmetries into the model, such as the static misalignment of the optics, which is the main coupling channel of the beam jitter to the gravitational wave signal.\\
The calculation method presented here has been validated in Virgo, by measuring accurately the input jitter as well as the transfer function of the coupling to the dark fringe \cite{VIR-0331A-11}.

The beam jitter requirements at the ITF input are shown in Figure~\ref{fig_jitter_requirements}. It shows stringent levels of a few tens of pm/sqrt(Hz) for shifts and a few frad/sqrt(Hz) for angles at frequencies of about 200 Hz.

In order to then evaluate these requirements at the input of the IMC, the filtering effect of the IMC itself and the shaping optics effect have to be taken into account.
Indeed these stringent requirements are relaxed by the use of the IMC filtering cavity~\cite{cav}. For a linear cavity the transverse vertical and horizontal modes have the same resonance frequency. However, for a ring cavity, this is not true. If the cavity has an odd number of mirrors, as is the case for the Advanced Virgo Input Mode cleaner, the modes TEMmn with an odd mode number relative to the ring plane (m = odd) are non-degenerate with respect to the modes TEMnm having the same mode number relative to the plane perpendicular to the ring \cite{Barsu98}. Thus, vertical and horizontal misalignment modes are filtered in a different way. In consequence, the vertical beam jitter is filtered by the Advanced Virgo IMC cavity by a factor of $\gamma_v = 340$ and the horizontal by a factor of $\gamma_h = 670$. \\
The shaping optics effect can be taken into account by using the ABCD matrix formalism. 
Considering that
\begin{equation}
\left[\begin{array}{cc} A & B \\ C & D \\ \end{array} \right] =  \left[\begin{array}{cc} 14.36 & -171.05 \\ -0.2 & 2.46 \\ \end{array} \right]
\label{eq:abcd}
\end{equation}
is the ABCD matrix of the shaping optics~\cite{TDR}, the lateral and angular  
beam jitter at the IMC input, x$_{IMC}$ and $\theta_{IMC}$, can be calculated starting from the jitter at the ITF input, x$_{ITF}$ and $\theta_{ITF}$, as:
\begin{eqnarray}
\label{eq1}
     x_{IMC}=\sqrt{|A' x_{ITF}|^2+|B' \theta_{ITF}|^2}\cdot \gamma_h \\
     y_{IMC}=\sqrt{|A' x_{ITF}|^2+|B' \theta_{ITF}|^2}\cdot \gamma_v 
\end{eqnarray}

\begin{eqnarray}
     {\theta_y}_{IMC}=\sqrt{|C' x_{ITF}|^2+|D' \theta_{ITF}|^2}\cdot \gamma_h \\
     {\theta_x}_{IMC}=\sqrt{|C' x_{ITF}|^2+|D' \theta_{ITF}|^2}\cdot \gamma_v 
\label{eq2}
\end{eqnarray}

where $x_{IMC}$ and $y_{IMC}$ are the horizontal and vertical shifts, ${\theta_y}_{IMC}$ and ${\theta_x}_{IMC}$ are the tilts in the yaw and pitch directions of the beam at the IMC input. Moreover, the parameters of the optical system $A' B' C' D'$ have been obtained by the inversion of the ABCD matrix of Equation~\ref{eq:abcd}.

\begin{figure}[h]
\begin{center}
\includegraphics[width=0.5\textwidth,clip]{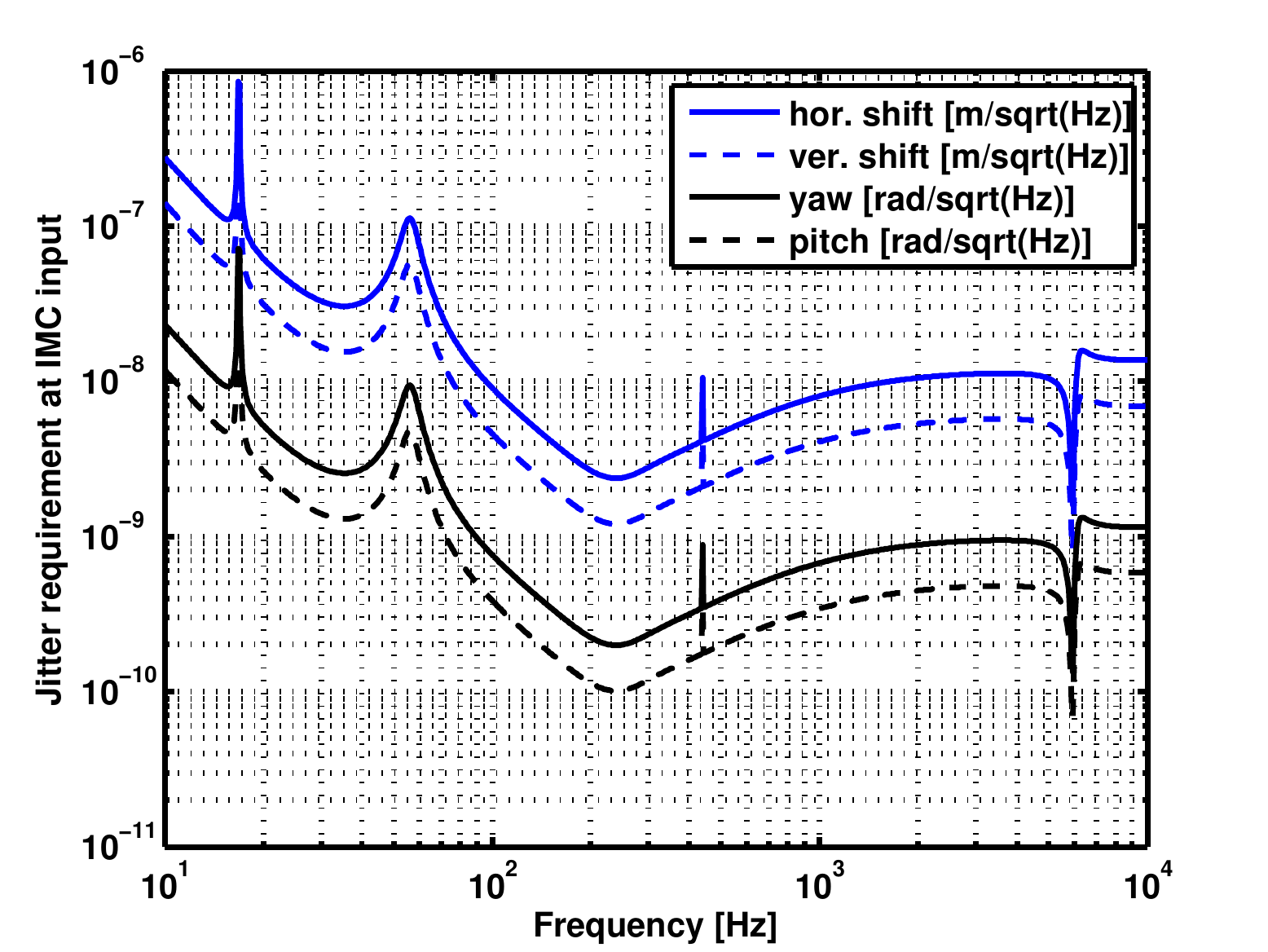}
\caption{\label{fig_jitter_requirements_imc} Advanced Virgo Jitter requirements at the input of the IMC (Input Mode Cleaner) for the shift and tilt d.o.f.s, blue and black curves respectively. The requirements for the horizontal and vertical directions are different due to the fact that the filtering of the IMC triangular cavity is different in the two directions. }
\label{default}
\end{center}
\end{figure}
Figure~\ref{fig_jitter_requirements_imc} shows the beam jitter requirements at the level of the IMC input. 
It can be observed that pointing noise requirement in the detector bandwidth is mitigated by more than one order of magnitude for the shift direction and of about four orders of magnitude for the tilt direction at the IMC input with respect to what has been computed for the ITF input.

\subsection{Input Beam Jitter low frequency accuracy requirements.}
\label{sec:accreq}

\begin{figure}[h]

    \centering
    \includegraphics[width=0.5\textwidth,clip]{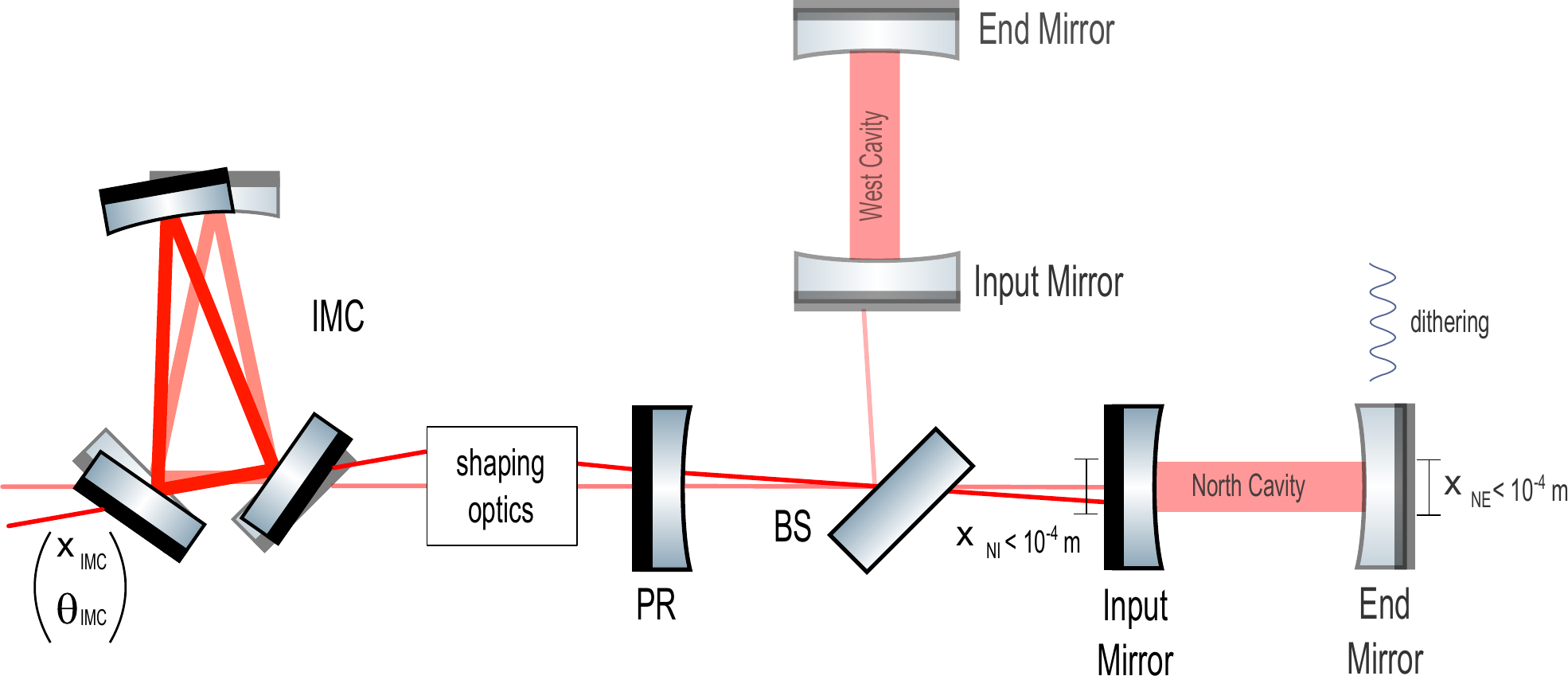}
    \caption{Scheme of the effect of a input beam shift and tilt ($x$ and $\theta$) at the entrance of the Input Mode Cleaner triangular cavity. The beam will pass trough the IMC unperturbed thanks to the IMC Automatic Alignment reaching the Input mirrors tilted and shifted after passing trough the shaping optics. The requirements of the input beam shift and tilt have been set in order to have a maximum displacement of the beam entering in the North cavity of 10$^{-4}$m.}
    \label{LFreq}
\end{figure}

Low frequency fluctuations of the beam alignment at the ITF input can impact lock accuracy and therefore limit the sensitivity \cite{vsr1}.
A tilted input beam in the ITF introduces a mismatch with the cavity mode. This mismatch is then compensated by the angular control of the arm cavity modes, the Automatic Alignment control system~\cite{TDR}, which acts on the cavity mirrors to re-align the cavity axis on the input beam.
This misalignment of the cavity axis unfortunately leads to a displacement of the beam on the cavity mirrors with respect to their center of rotation thereby increasing the longitudinal/angular noise coupling and thus the contribution of the angular control noise to the sensitivity. The Evaluation of the angular sensing noise reveals that the maximum allowable displacement on the cavity mirrors, both on the Input and End mirrors, to achieve the Advanced Virgo sensitivity requirements is 10$^{-4}$m~\cite{AAdes}.
For the end mirrors the centering of the beam is obtained thanks to the implementation of a local control strategy, called \textit{dithering} already implemented in the Virgo configuration~\cite{vsr2}, which maintains the beam centered with respect to the End mirrors center of rotation within the accuracy requirement. The net requirement on ITF input beam alignment then consists of controlling the shift on the input mirrors to 10$^{-4}$m.\\
In order to compute the accuracy requirement the effects of the IMC and of the shaping optics have  to be taken into account in the beam propagation.
The beam exiting from the laser system is used as a reference for the alignment of the wole interferometer and of the Input Mode Cleaner (IMC).
The low frequency misalignments of the IMC input beam are therefore compensated by  a misalignment of the IMC mirrors that simply follows the direction of the input beam (see Figure \ref{LFreq}) with a control bandwidth of $\sim$ 1 Hz \cite{3kmVirgo}.
The low frequency fluctuations of the beam alignment at the output of the laser system therefore remains almost unchanged passing through the IMC cavity.
Calculating requirements of low frequency misalignments at the IMC input nevertheless needs to take into account the effect of the shaping optics placed between the IMC and ITF. If $\left[   \begin{array}{c} x_{IMC} \\ \theta_{IMC} \\ \end{array} \right]$ represents the misaligned beam at the IMC input, the beam displacement at the North cavity input mirror can be computed as:
\begin{equation}\label{spec}
x_{IM} = \left[ 1 ~~L \right] \cdot \left[
\begin{array}{cc}
 A & B \\
 C & D
\end{array}
\right]
\cdot \left[
\begin{array}{cc}
 1 & d \\
 0 & 1
\end{array}
\right] \cdot \left[
\begin{array}{c}
 x_{IMC} \\
 \theta_{IMC}
\end{array}
\right]  <10^{-4}\mathrm{m}
\end{equation}
where $d$ is the distance between the first two mirrors of the IMC, the ABCD matrix is given by  Equation~\ref{eq:abcd} and L is the distance between the PR mirror and the input mirror (IM).
The requirements on the beam shift, $x_{IMC}$, and beam tilt, $\theta_{IMC}$, are then~\cite{notareq}:
\begin{equation}\label{spec_r_t}
    \begin{array}{ccc}
      x_{IMC} & < & 4.2\mu\mathrm{m}\\
      \theta_{IMC} & < & 0.35 \mu\mathrm{rad}
    \end{array}
\end{equation}

These requirements are hard to achieve as large pointing fluctuations are expected at low frequency due to temperature variations and air flux. \\

In order to evaluate and to ensure that the jitter of the beam at the IMC input port is compatible with requirements calculated in the previous sections~\ref{sec:highfrequreq} and \ref{sec:accreq}, a system, to monitor shifts and tilts in the frequency band 10~Hz - 10~kHz and to mitigate the jitter at frequencies below 10~Hz, has to be developed.

\section{Beam Pointing Control system design and experimental setup}\label{sec:des}

\begin{figure}[htp]
    \centering
    \includegraphics[width=0.5\textwidth,clip]{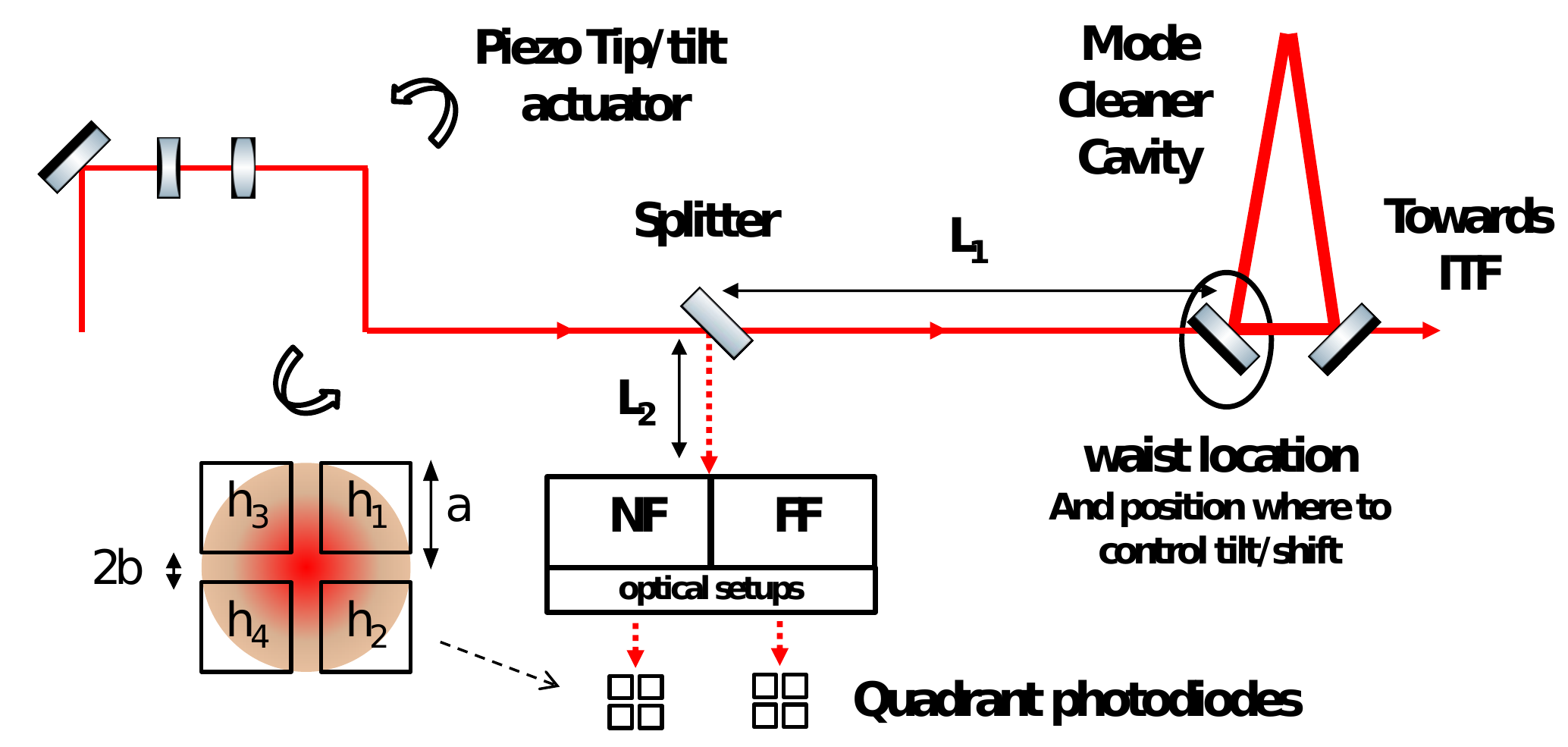}
    \caption{Principle of the Beam Pointing Control.}
    \label{BPC_principle}
\end{figure}

The Beam Pointing Control system has been developed to reduce the input beam tilt and shift at frequencies below 10Hz. The shift and tilt of the beam are sensed by two quadrant photo-diodes placed at the input port of the IMC and the beam is steered to the IMC input by using a system of Piezo actuators, as is shown in Figure~\ref{BPC_principle}.  

The beam going from the laser is mode-matched onto the Input Mode Cleaner (IMC) cavity using a telescope. A partially reflecting mirror is used on the beam path to obtain a pick-off before the IMC and to sense the shifts and the tilts at the input of the cavity (named hereafter Near field (NF) and Far Field (FF) respectively). 
\subsection{Sensing: analytical computation and experimental setup}

The beam tilt and shift are converted by the optical setup into displacements sensed by two quadrant photo-diodes placed in a way as to get 90~deg of Gouy phase difference between them. 
The quadrant signals are then used in a feed-back loop using two tip/tilt piezo mirrors as actuators.
\subsubsection{Analytical computation}
In this section the sensitivity of the NF and FF sensing setup will be calculated. 
The first step is to evaluate the quadrant sensitivity to the beam displacement.
Considering a beam, with a beam radius $w$, which impinges the quadrant diode, the intensity on the sensor can be written as:
\begin{equation}\label{Iexpression}
    I(x,y)=\frac{2I_0}{\pi  w^2}e^{-\frac{2 (x+y)^2}{w^2}}
\end{equation}
Considering then that the quadrant is composed of four distinct square zones of size $a$, separated by a gap $2b$ (see Figure~\ref{BPC_principle}), the quadrant horizontal output signal for a beam misalignment ($x_{qd}$,$y_{qd}$) in the two directions can be written as:
\begin{eqnarray*}
    S(x_{qd},y_{qd})&=&\alpha I\ast (h_1(x_{qd},y_{qd})+h_2(x_{qd},y_{qd})\\
   &&-h_3(x_{qd},y_{qd})-h_4(x_{qd},y_{qd}))
\end{eqnarray*}
where $\alpha$ is the quadrant photo-diode responsivity and the function h$_i$ describes the response of the different zones of the sensor which can be expressed using the \textit{heaviside} function $\Theta$:
\begin{eqnarray*}
  h_1(x,y) &=& \Theta (a-x) \Theta (a-y) \Theta (x-b) \Theta (y-b) \\
  h_2(x,y) &=& \Theta (a-x) \Theta (a+y) \Theta (x-b) \Theta (-b-y) \\
  h_3(x,y) &=& \Theta (a+x) \Theta (a+y) \Theta (-b-x) \Theta (-b-y)\\
  h_4(x,y)&=& \Theta (a+x) \Theta (a-y) \Theta (-b-x) \Theta (y-b).
\end{eqnarray*}
After convolution using the Equation~\ref{Iexpression}, it can be found that:
\begin{equation}
    S(x_{qd},y_{qd})=\frac{\alpha I_0}{4}U_{+}(x_{qd})U_{-}(y_{qd}).
\end{equation}
where
\begin{eqnarray*}
  U_{\pm}(x_{qd}) &=& \text{erf}\left(\frac{\sqrt{2} (a-x_{qd})}{w}\right)\mp\text{erf}\left(\frac{\sqrt{2} (a+x_{qd})}{w}\right) \\
       &\pm&  \text{erf}\left(\frac{\sqrt{2} (b+x_{qd})}{w}\right) -\text{erf}\left(\frac{\sqrt{2} (b-x_{qd})}{w}\right)
 \end{eqnarray*}
In the following, for simplicity, only the horizontal response $S(x_{qd})$ of the sensor for a perfectly vertically aligned beam will be considered:
\begin{equation}
   S(x_{qd},0)=S(x_{qd})=\frac{\alpha I_0}{4}U_{+}(x_{qd})U_{-}(0)
\end{equation}
with
\begin{equation}
    U_{-}(0)=2\text{erf}\left(\frac{\sqrt{2} a}{w}\right)-2\text{erf}\left(\frac{\sqrt{2} b}{w}\right)
\end{equation}
For $x_{qd}\ll w$ we obtain:
\begin{equation}
    U_{+}(x_{qd})\approx\frac{4 \sqrt{\frac{2}{\pi }} }{w} \left(e^{-\frac{2 a^2}{w^2}}-e^{-\frac{2 b^2}{w^2}}\right)x_{qd}
\end{equation}
and the response of the sensor can be linearized as:
\begin{equation}
    S_h(x_{qd})=S x_{qd} 
    \label{eq:9}
\end{equation}
with
\begin{equation}
    S=\frac{\alpha I_0 \sqrt{\frac{2}{\pi }} }{w} U_{-}(0)\left(e^{-\frac{2 a^2}{w^2}}-e^{-\frac{2 b^2}{w^2}}\right).
\end{equation}
\begin{figure}[htp]
    \centering
    \includegraphics[width=0.5\textwidth,clip]{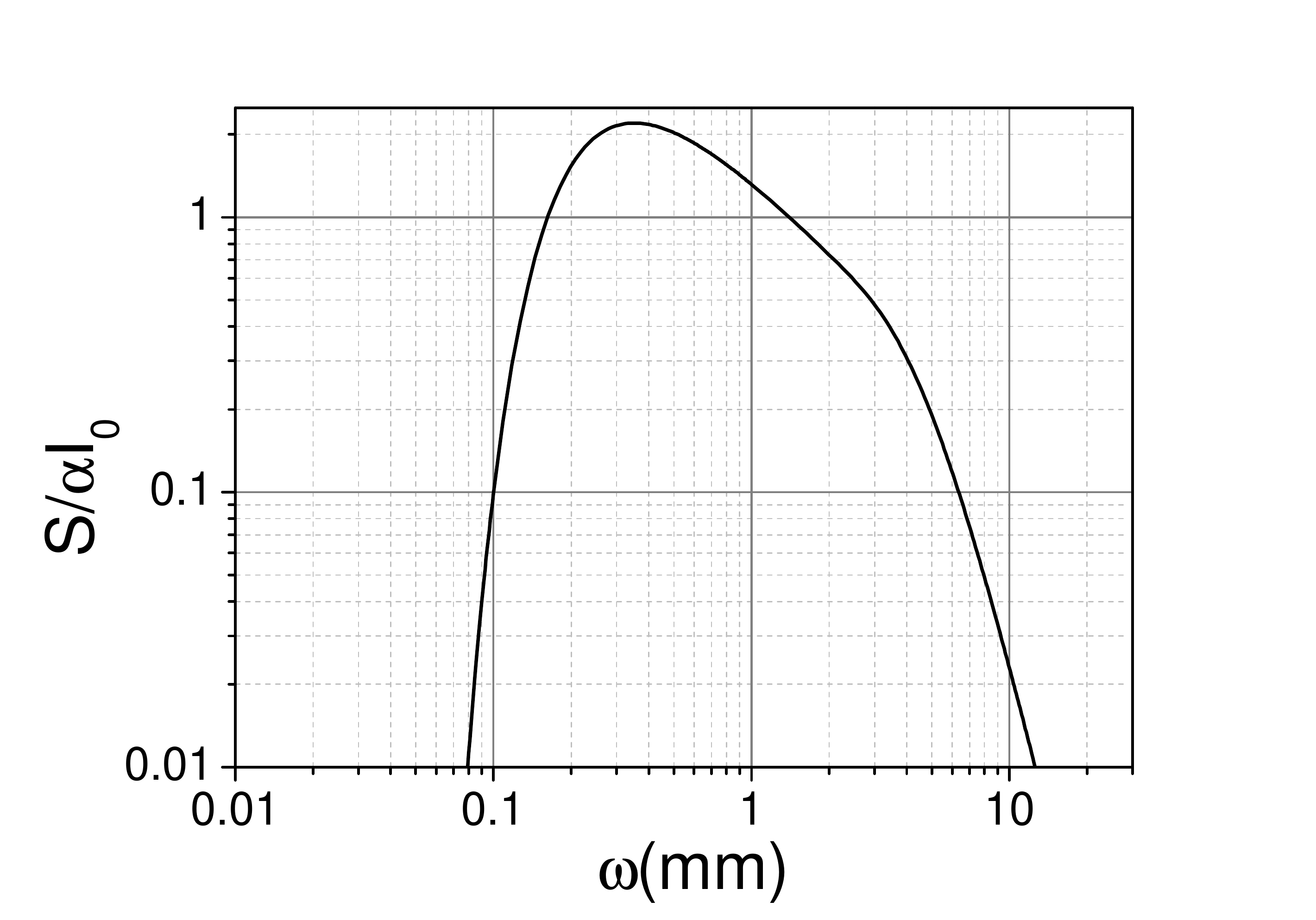}
    \caption{$S/\alpha I_0$ behavior as a function of $w$ for a detector gap of $2b=200$~$\mu$m and size $a=3.9$~mm}
    \label{plotfactor}
\end{figure}
Figure~\ref{plotfactor} shows the variation of $S/\alpha I_0$ as a function of $w$ for a detector gap of $2b=200$~$\mu$m and size $a=3.9$~mm. As expected the detector sensitivity drops to zero when $w\ll b$ and when $w\gg a$.
For a beam radius much larger than the gap, $w\gg b$, and much smaller than the sensor size, $w\ll a$, the normalized signal is:
\begin{equation}
  S_N = \frac{S}{\alpha I_0}\approx\frac{2 \sqrt{\frac{2}{\pi }}}{w}
  \label{eq:11}
\end{equation}
The second step is to design the optical configuration of this setup. This optical system should be designed in order to detect the pure shift of the beam on the IMC input on the NF quadrant diode and the pure tilt on the FF quadrant diode~\cite{cleva}.

The beam shift and tilt at the input of the IMC is described by the vector $\left[ \begin{array}{c} x_{iMC} \\ \theta_{IMC} \end{array} \right]$, but in the following a more general notation will be used, $\left[ \begin{array}{c} x_{in} \\ \theta_{in} \end{array} \right]$ which defines a general input beam for which the displacement must be sensed. When placing a lens at a distance $-d=-(L_1-L_2)$ from the FF and NF position, the displacement of the impinging beam on the quadrant will be:
\begin{equation}
\left[ \begin{array}{c} x_{qd} \\ \theta_{qd} \end{array} \right]=
\left[
\begin{array}{cc}
 1 & d' \\
 0 & 1
\end{array}
\right]
\left[
\begin{array}{cc}
 1 & 0 \\
 -\frac{1}{f} & 1
\end{array}
\right]
   \left[
\begin{array}{cc}
 1 & -d \\
 0 & 1
\end{array}
\right]\left[ \begin{array}{c} x_{in} \\ \theta_{in} \end{array} \right].
\end{equation}
The displacement on the sensor is then:
\begin{equation}
x_{qd}= x_{in} \left[1-\frac{d'}{f}\right]+\theta_{in}  \left[d'-d \left(1-\frac{d'}{f}\right)\right] \\
\end{equation}
Then the FF can be sensed by placing a quadrant at $d'=f$ and the NF by placing a sensor at $d'=\frac{df}{d+f}$.
Since the quadrant sensitivity drops to zero for small impinging beams, for which $w^2 \ll b^2$, the one-lens setup cannot be used for the the FF since it requires placing the quadrant in the lens focus.\\
For the NF the displacement $x_{qd}$ measured by the sensor is independent of $\theta_{in}$ and becomes:
\begin{equation}
  x_{qd}=K_{NF} x_{in}
  \label{eq:14}
\end{equation}
where $K_{NF}$ is the displacement amplification factor:
\begin{equation}
  K_{NF}=\frac{f}{d + f}.
\end{equation}
The quadrant signal, by using the eq.~\ref{eq:9}~\ref{eq:11}~\ref{eq:14}, can be written as:
\begin{equation}
     S_{NF}(x_{in})=2 \sqrt{\frac{2}{\pi }}\frac{K_{NF}}{w_{NF}}x_{in}
\end{equation}
where $w_{NF}$ is the beam radius at the position of the quadrant sensor. The beam radius as a function of the distance z from the lens can be written as:
\begin{equation}\label{wNFz}
w^{'2}(z)=\frac{\lambda}{\pi} z'_R\left[ 1+\left(\frac{z-z'_0}{z'_R}\right)^2\right]
\end{equation}
where $z'_0$ and $z'_R$ are respectively the waist position and the Rayleigh range of the beam after propagation though the lens. Using the ABCD matrix propagation, it can be found:
\begin{equation}
\left\{\begin{array}{ccc}
  z'_0 & = & \frac{z_R^2(K_{NF}d/f^2-1/f)}{1/K_{NF}^2+z_R^2/f^2} \\
  z'_R & = & \frac{z_R}{1/K_{NF}^2+z_R^2/f^2}
\end{array}
\right.
\end{equation}
$w_{NF}$ can then be calculated from eq. \ref{wNFz}:
\begin{equation}
w_{NF}^2=w^{'2}(K_{NF}d)=\frac{\lambda}{\pi} z'_R\left[ 1+\frac{z_R^2}{f^2}(\frac{K_{NF}d}{f}-1)^2\right].
\end{equation}
With $(\frac{K_{NF}d}{f}-1)^2=K_{NF}^2$, it becomes:
\begin{equation}
w_{NF}^2=\frac{\lambda}{\pi} K_{NF}^2 z_R=K_{NF}^2 w_0^2
\end{equation}
and the quadrant signal can be written as:
\begin{equation}
     S_{\scriptstyle NF}(x_{in})=2 \sqrt{\frac{2}{\pi }}\frac{x_{in}}{w_0}
\label{eq:sensNF}
\end{equation}
The sensitivity of NF sensing with a quadrant does not depend on the optical setup but only on the waist size of the monitored beam.
For the FF, a telescope formed by two lenses of focal length f$_1$ and f$_2$ separated by a distance d$_{12}$ is used to shape the beam on the quadrant diode, i.e. to obtain the beam size which maximizes the sensitivity for the given quadrant photo-diode parameters ($a$ and $b$) as it is shown in Figure~\ref{plotfactor}.
\begin{widetext}
\begin{equation}
\left[ \begin{array}{c} x_{qd} \\ \theta_{qd}  \end{array} \right]=
\left[
\begin{array}{cc}
 1 & d' \\
 0 & 1
\end{array}
\right]
\left[
\begin{array}{cc}
 1 & 0 \\
 -\frac{1}{f_2} & 1
\end{array}
\right]
\left[
\begin{array}{cc}
 1 & d_{12} \\
 0 & 1
\end{array}
\right]
\left[
\begin{array}{cc}
 1 & 0 \\
 -\frac{1}{f_1} & 1
\end{array}
\right]
   \left[
\begin{array}{cc}
 1 & -d \\
 0 & 1
\end{array}
\right]
\left[ \begin{array}{c} x_{in}  \\ \theta_{in} \end{array} \right]
\end{equation}
The displacement on the sensor is then:
\begin{equation}
   x_{qd}=x_{in} \left(-\frac{d'}{f_2}-\frac{d'+d_{12} \left(1-\frac{d'}{f_2}\right)}{f_1}+1\right)+\theta_{in}  \left(d'+d_{12} \left(1-\frac{d'}{f_2}\right)+d
   \left(-\frac{d'}{f_2}-\frac{d'+d_{12} \left(1-\frac{d'}{f_2}\right)}{f_1}+1\right)\right)
\end{equation}
\end{widetext}

For a quadrant placed at a distance $d'=\frac{f_2 \left(d_{12}-f_1\right)}{d_{12}-f_1-f_2}$, the displacement $x_{qd}$ measured by the sensor is independent of $x_{in}$ and becomes:

\begin{equation}
  x_{qd}=K_{FF}\theta_{in}
\end{equation}
with
\begin{equation}
K_{FF}=\frac{f_1 f_2}{f_1+f_2-d_{12}}
\end{equation}
and the quadrant signal can be written as:
\begin{equation}
     S_{FF}(\theta_{in})=2 \sqrt{\frac{2}{\pi }}\frac{K_{FF}}{w_{FF}}\theta_{in}
\end{equation}
where $w_{FF}$ is the beam radius at the position of the quadrant sensor. In the same way as for NF, the ABCD matrix propagation has been used to calculate $w_{FF}$:
\begin{equation}
w_{FF}^2=\frac{\lambda}{\pi} K_{FF}^2 \frac{1}{z_R}=\left(\frac{\lambda}{\pi}\right)^2\frac{K_{FF}^2}{w_0^2}
\end{equation}
and the quadrant signal can be written as:
\begin{equation}
     S_{FF}(\theta_{in})=2 \sqrt{\frac{2}{\pi }}\frac{\theta_{in}}{\lambda/(\pi w_0)}
\label{eq:sensFF}
\end{equation}

Similarly to what was obtained for the NF, the FF quadrant sensitivity is independent of the setup and only depends on the divergence $\lambda/(\pi w_0)$ of the monitored beam.

\subsubsection{Experimental setup}

\begin{figure}[htp]
    \centering
    \includegraphics[width=0.4\textwidth,clip]{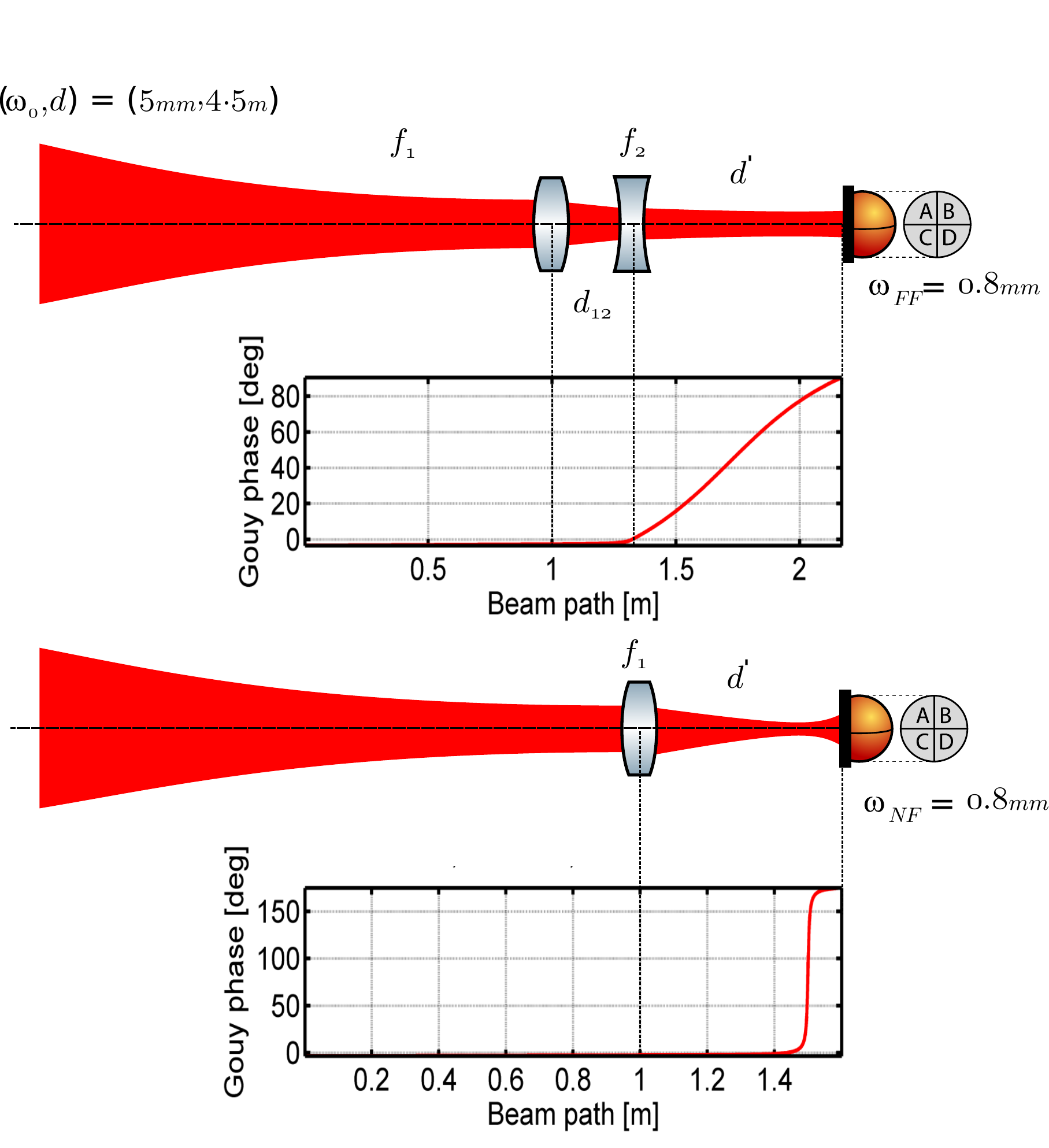}
    \caption{Top picture: BPC Far Field setup showing the beam and Gouy phase propagation; Bottom picture: BPC Near Field setup showing the beam and Gouy phase propagation.}
    \label{fig:Telescopes_BPC}
\end{figure}

The tilt and the shift of the beam at the input of the IMC cavity is monitored by the two quadrant photo-diodes, NF and FF, placed at a Gouy phase difference of $\Delta\phi_{Gouy} = \phi_{NF} - \phi_{FF} = 90$ deg. 
The two quadrant photo-diodes, $S - 078 - QD$ provided by EOS Inc.$^{TM}$ \cite{eos_inc}, are composed of four separate sensitive elements with an active area of 7.8~mm and they are separated by a gap of less than 200~$\mu$m. 
Considering the gap and the dimension of the photosensitive zone, the beam impinging on the quadrant photo-diode should have a radius between 800 $\mu$m and 1mm.\\
The sensing is formed by two different optical paths: the NF sensing path, to sense the pure shift of the beam entering the IMC, and the FF sensing path to sense the pure tilt, as it has been described in section~\ref{sec:des}.\\
The Near Field sensing path is composed of a single lens, with a focal length of $f_1 = 0.5$ m and a quadrant at a distance $d' = 0.6$ m from the lens, such as to be in the image plane, as shown in Figure~\ref{fig:Telescopes_BPC}. 
The Far Field sensing path is composed of a system of two lenses with focal length $f_1 = 0.356$ m and $f_2 = - 0.0387$ m. The distance between them is optimized to $d_{12} = 0.319$ m in order to obtain a telescope which is equivalent to a converging lens with an effective focal length of $f_{eq} \simeq 8$~m. 
In order to place the quadrant diode in the focal plane, to be sensitive only to tilts of the input beam, the distance between the quadrant and the second lens of the telescope ($l_2$) has been set to $d' = 0.85$ m yielding a Gouy phase shift between the two sensing photo-diodes of 90~deg, as shown in Figure~\ref{fig:Telescopes_BPC}.

\begin{figure}[htp]
    \centering
    \includegraphics[width=0.5\textwidth,clip]{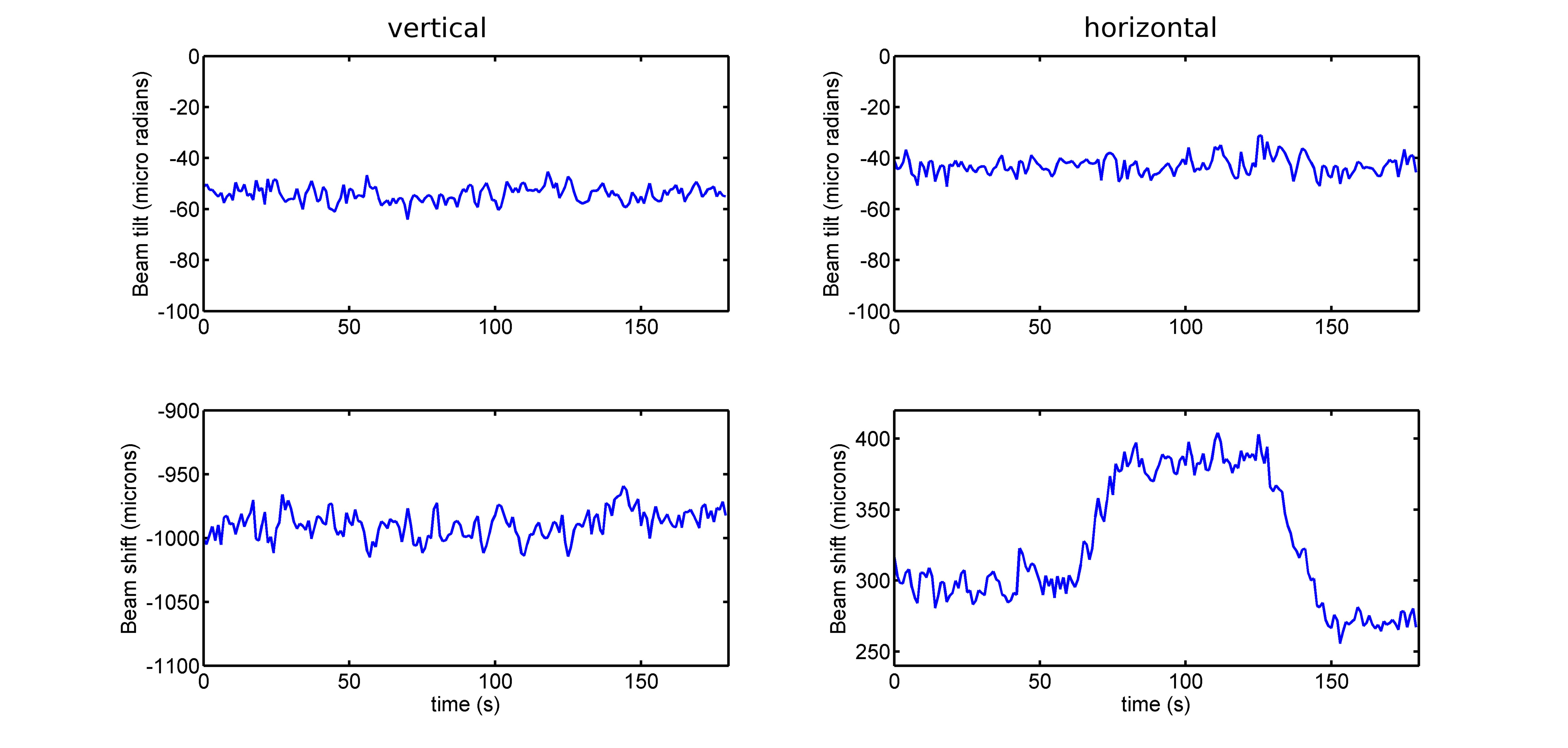}
    \caption{Shift measurement: BPC FF quadrant outputs (top); BPC NF quadrant outputs (bottom).}
    \label{fig:shift}
\end{figure}
The two quadrants have been placed 90~deg apart in order to perfectly decouple the two degrees of freedom; the tilt and the shift of the beam.
 The perfect decoupling of the sensing has been proved by purely shifting the beam in the horizontal direction by inserting and rotating a glass window with anti-reflective coating at 1.064 $\mu$m laser wavelength in the beam path. As it is shown in Figure~\ref{fig:shift}, the effect is visible only in the horizontal signal which detects the beam shift (on the Near Field quadrant).\\
The quadrant signals will be used at low frequency as error signals for the beam pointing control, and at high frequency, in the detection frequency band 10Hz-10kHz, to monitor the beam jitter.

\subsection{Actuation experimental setup}

\begin{figure}[htp]
    \centering
    \includegraphics[width=0.5\textwidth,clip]{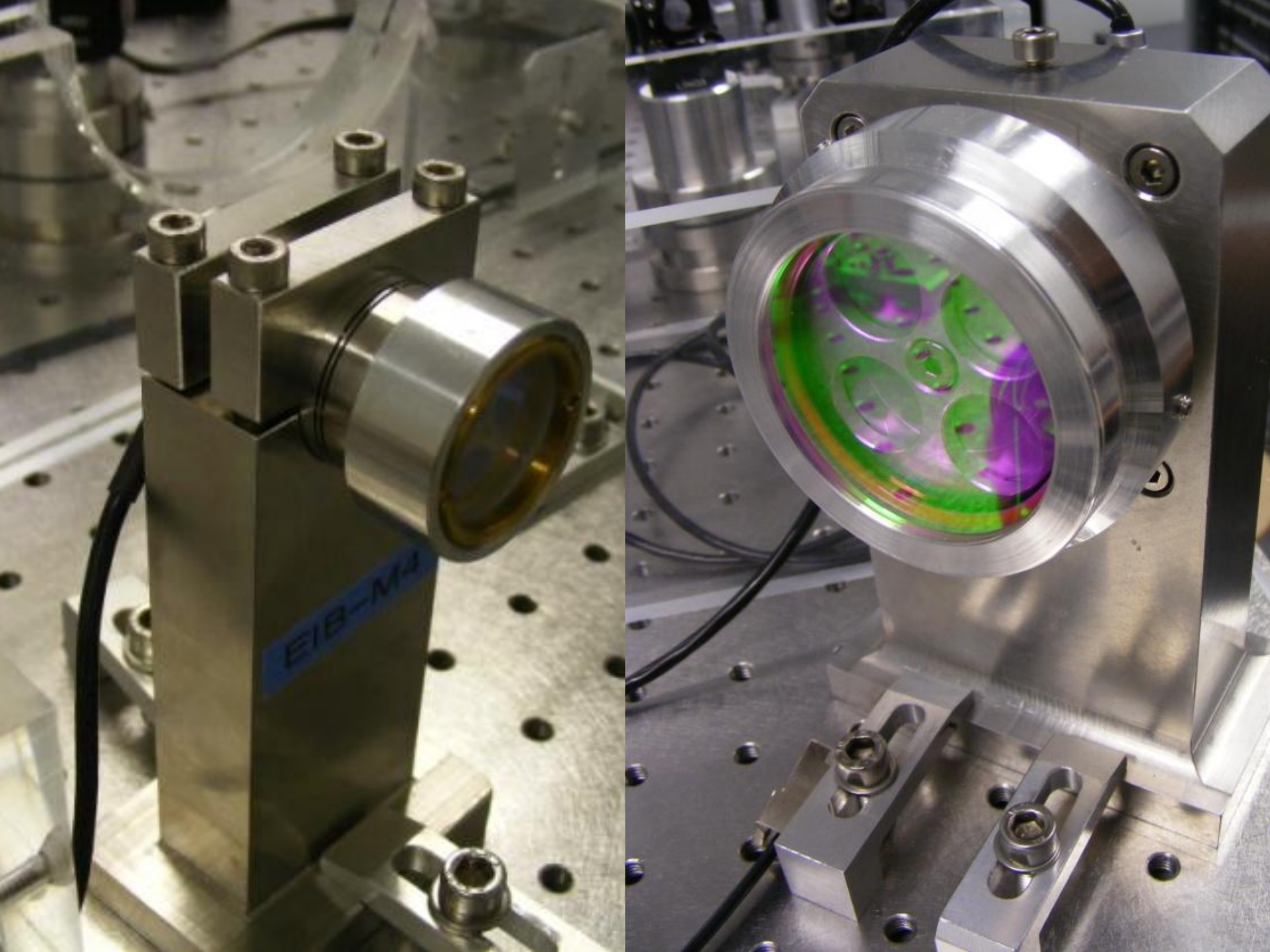}
    \caption{Mirror mounted on the piezoelectric actuator PI S330 on the left; Mirror mounted on the piezoelectric actuator PI S340 on the right.}
    \label{fig:piezoview}
\end{figure}

In order to adjust the position of the beam on the first mirror of the Input Mode Cleaner Cavity and reduce beam jitter noise at low frequency, two steering mirrors are used. In particular, they are composed of two systems in which mirrors and piezoelectric actuators are fixed in the same mechanical mount. 
Two different models of tip/tilt piezoelectric actuators have been used: the S330 and S340 from Physik instrumente~\cite{piezo} shown in Fig. \ref{fig:piezoview}. 

\section{BP Control System: calibration procedures and control scheme}

The Beam Pointing Control system is a feed-back control loop aiming at suppressing the input beam displacement at the level of the IMC first mirror, in terms of tilt and shift, in the frequency range below 10~Hz.
The implementation of the control requires introducing well defined calibration procedures for both the actuation and the sensing.

\subsection{Actuation calibration procedure}

\begin{figure}[htp]
    \centering
    \includegraphics[width=0.5\textwidth,clip]{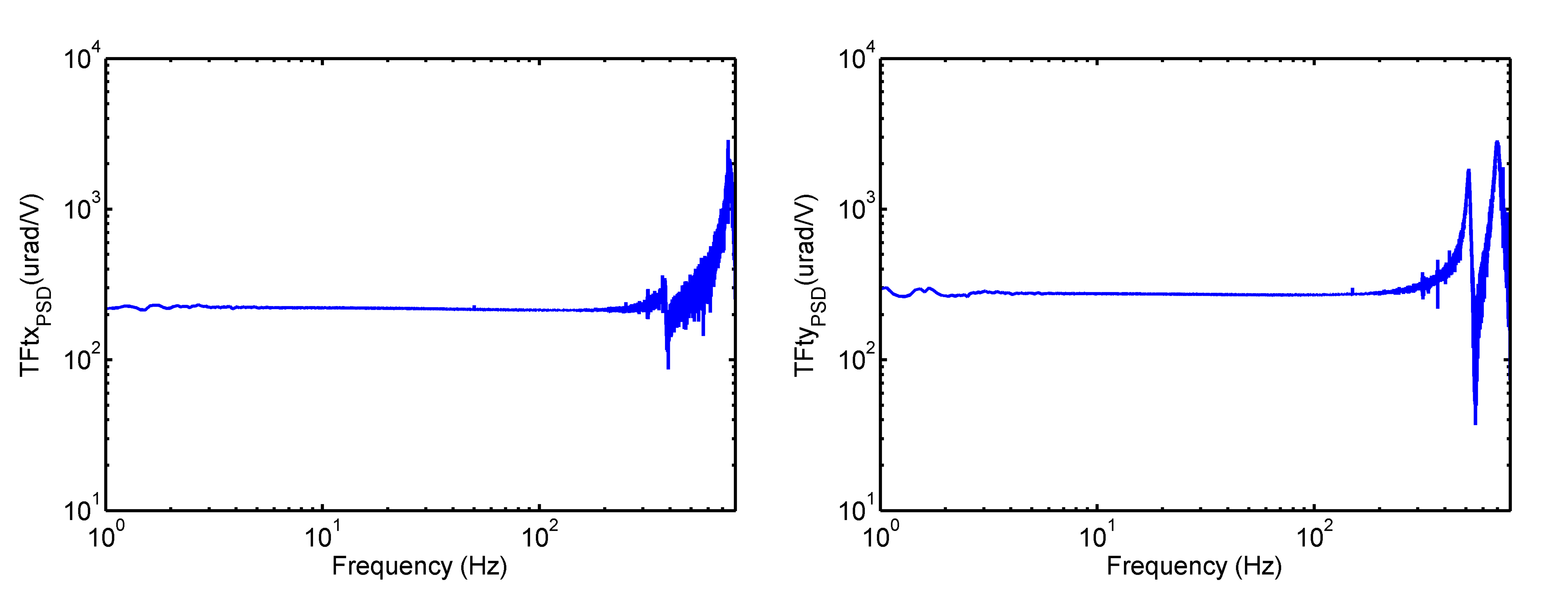}
    \caption{Transfer function of pitch and yaw direction for the PI S340 piezo actuator ($Pzt_{2}$).}
    \label{fig:calibpiezo}
\end{figure}

The actuation system has been calibrated with a very basic setup using a Position Sensor Device (PSD) mounted on a micro-metric translation stage, from Optosigma, placed in the focal plane of a lens having the focal length $f=150$mm. \\ 
The PSD will then only be sensitive to tilts and it can be easily used to calibrate the two piezo actuator ($Pzt_{1/2}$). 
The result of this calibration is depicted in Figure~\ref{fig:calibpiezo} and shows a flat Transfer Function between the voltage applied to the second piezo and the angular displacement induced on the PSD up to a few hundreds of Hertz.

\subsection{Sensing calibration procedure}
 
 \begin{figure}[htp]
    \centering
    \includegraphics[width=0.5\textwidth,clip]{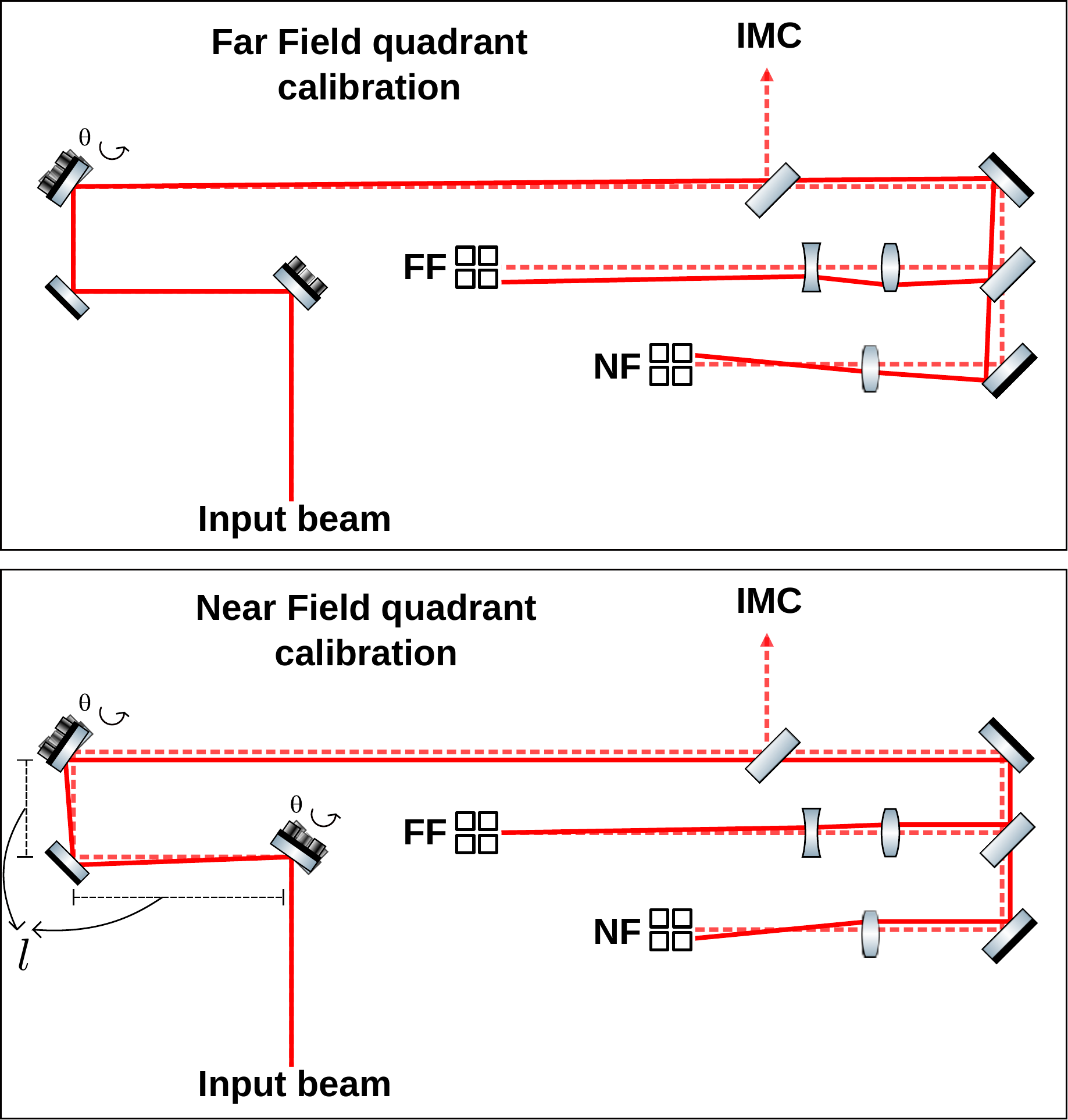}
    \caption{Sensing calibration procedure of the Far Field quadrant, top plot, and the Near Field quadrant, bottom plot.}
    \label{fig:calib_sensing}
\end{figure}

The quadrant diodes have been calibrated applying known shift and tilt of the beam by using the previously calibrated piezo actuators, as shown in Figure~\ref{fig:calib_sensing}.\\
The FF quadrant calibration is performed by injecting a calibrated tilt ($\theta$) using the calibrated piezo actuator ($Pzt_{2}$) and measuring the consequent quadrant signal as output. The NF quadrant is calibrated applying the same amount of tilt ($\theta$) on both piezo actuators, obtaining a pure shift and no signal on the Far Field quadrant taking into account also the distance $l=1.392$~m between the two piezo actuators.
The calibration procedure of the sensing yields a FF normalized signal of  ${{S_{FFtx}}\vert}_{meas}$=14045~[rad$^{-1}$] for the vertical d.o.f. and ${{S_{FFty}}\vert}_{meas}$=18182~[rad$^{-1}$] for the horizontal d.o.f.. 
The calibration in the two directions is slightly different due to the fact that the beam size in the two directions is not equal ($w_{FFy}=800\mu m$ in the vertical direction and $w_{FFx}=627\mu m$ in the horizontal direction).
These results are in good agreement with the expected results ${{S_{FFtx}}_\vert}_{teo}$=16166~[rad$^{-1}$] for the vertical d.o.f. and ${{S_{FFty}}\vert}_{teo}$=20626~[rad$^{-1}$] obtained from the eq.~\ref{eq:sensFF}.
In the case of the Near Field signal the measured value is ${{S_{NFx}}\vert}_{meas}$=492~[m$^{-1}$] for the horizontal direction and  ${{S_{NFy}}\vert}_{meas}$=431~[m$^{-1}$] for the vertical direction, to be compared with the theoretical result, using the eq.~\ref{eq:sensNF}, of ${{S_{NF}}\vert}_{teo}$=398~[m$^{-1}$] for a beam waist of $w_{NFx}=800\mu m$.
The good agreement between the measured values and the expected results in addition to the validation of the analytical computation could also allow to design the control using only the theoretical calibrations.

\subsection{Control scheme}

\begin{figure}[htp]
    \centering
    \includegraphics[width=0.45\textwidth,clip]{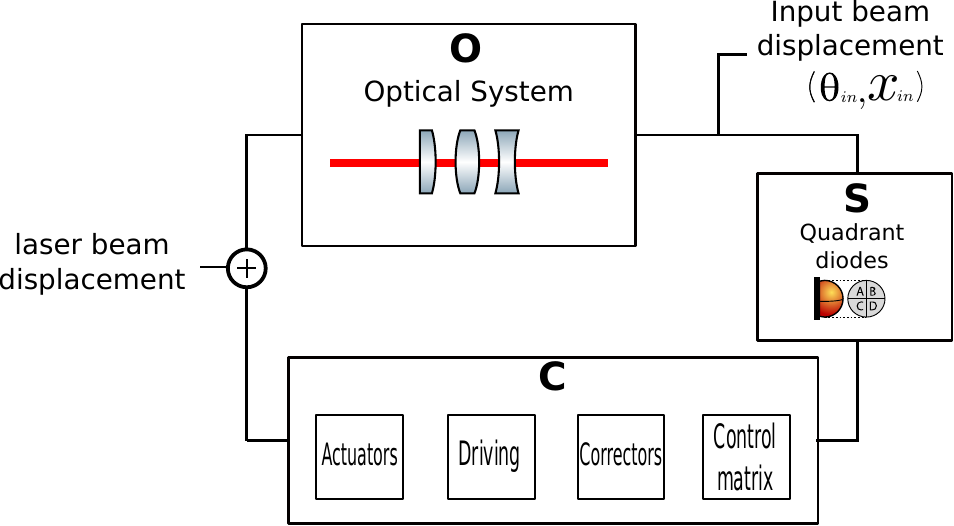}
    \caption{Block scheme for the Beam Pointing Control system.}
    \label{fig:loop}
\end{figure}

The Beam Pointing Control scheme, whose block diagram is shown in Figure~\ref{fig:loop}, is composed of the  the Optical system (O), the  Sensing part (S) and the Control part (C).\\
The \textit{Optical system} (O) represents the optical path response from the piezo actuator tilts to the input beam displacements.
The Sensing (S) is the opto-electronic part of the system consisting of the two quadrant diodes and the relative readout electronics to extract the information about the input beam displacements.
\begin{table}[htb]
 \centering 
 \begin{tabular}{c|c|c|c|c} 
 & \textbf{$FF_{hor}$} &  \textbf{$FF_{ver}$} &  \textbf{$NF_{hor}$} &  \textbf{$NF_{ver}$}\\
 \hline
 \textbf{$\theta_x$} & 0 & 71.2 & 0 & 0 \\
 \textbf{$\theta_y$} & -55 & 0 & 0 & 0 \\
 \textbf{$x$} & 0 & 0 & 2034 & 0 \\
 \textbf{$y$} & 0 & 0 & 0 & 2319 \\
 \end{tabular} 
\caption{\label{table:Sensing}\small{Control matrix in [$\mu$m] for the $x$ and $y$ direction and in [$\mu$rad] in the $\theta_x$ and $\theta_y$ direction. }} \end{table}
The \textit{Control} (C) is given by the Control Matrix, the Correctors, the Driving matrix and the Actuator transfer functions.
The control matrix provides the relation between the set of quadrant signals and the set of input beam displacements. It's entries are the low frequency limit of the transfer functions between the input beam displacements and the quadrant signals (see Table~\ref{table:Sensing}).
The BPC corrections are computed by filtering the error signals with properly designed control filters (correctors) which are sent then to the driving matrix to define the correction voltages to drive the piezo-electric actuators.
\begin{figure}[htp]
    \centering
    \includegraphics[width=0.5\textwidth,clip]{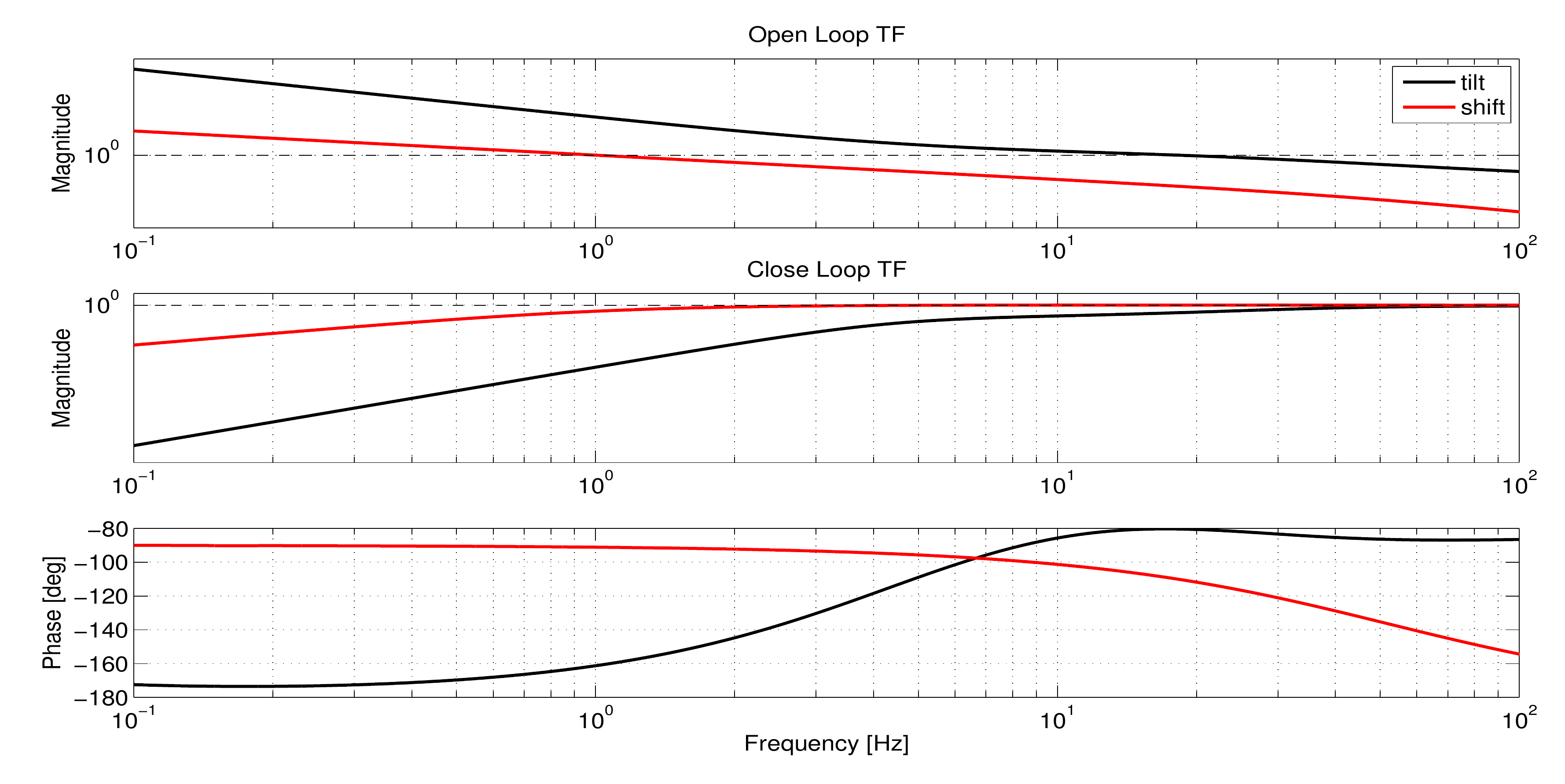}
    \caption{BPC open loop Transfer Function on the top plot, closed loop TF on the middle plot and the phase on the bottom plot for the shift and tilt control filters (red and black curves respectively). As it is shown in the transfer function the unity gain frequency of the two loops is $\sim$ 20~Hz for the tilt direction and $\sim$1~Hz for the shift direction.}
    \label{fig:OLCLfilt}
\end{figure}
The control filters for the shift and tilt are shown in Figure~\ref{fig:OLCLfilt}. The  unity gain frequency of the two loops is $\sim$ 20~Hz for the tilt direction (black curve) and $\sim$1~Hz for the shift direction (red curve).
The control filters, developed for the BPC control, are very simple and robust. From the open loop transfer function it can be seen that they demonstrated a large margin of improvement, the u.g.f. could be indeed increased to $\sim$100~Hz for both directions, which is not necessary for the purpose of this study, since the system is already compliant with the control accuracy and noise requirements, but it could be useful for other applications.\\
The acquisition of the error signals and the generation of the corrections is made trough a Digital Signal Processor (DSP) with a sampling rate of 10~kHz.

\section{BPC performance}
As discussed in the previous section the BPC system has a dual purpose. The first is to mitigate the jitter at low frequency in order to not degrade the overall ITF alignment (see section~\ref{sec:accreq}). The second is to monitor the beam jitter at the level of the IMC.
In this section the performance of the experimental apparatus will be compared with the Advanced Virgo requirements.
\subsection{Low frequency performance: closed loop signals}

\begin{figure}[htp]
    \centering
    \includegraphics[width=0.5\textwidth,clip]{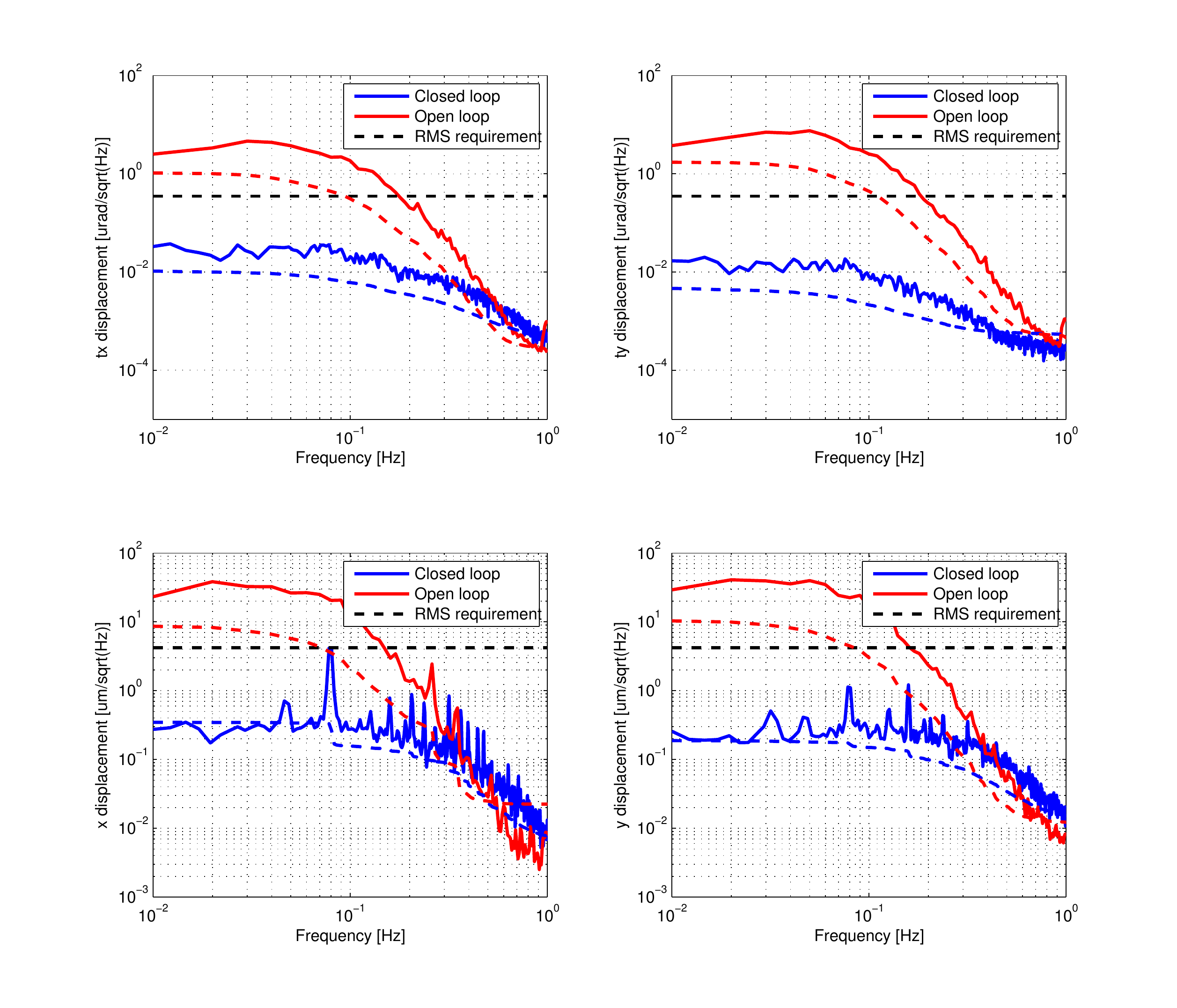}
    \caption{Beam shift and tilt spectra at the level of the input of the Mode Cleaner cavity in case of open and close loop (red and blue curve respectively) for all the directions. The open loop spectra have been computed with a statistical approach setting a confidence threshold of 95\%, which means that the beam displacement spectra will be below that curve for the 95\% of the time. The control accuracy RMS (the blue dash curve) fulfill the requirement (the dashed black curve) for all the four degrees of freedom. }
    \label{fig:OLCL}
\end{figure}
The low frequency performance of the control is shown in Figure~\ref{fig:OLCL}. The beam shift and tilt spectra at the level of the IMC in the case of open and closed control loop (red and blue curve respectively) for all the four d.o.f. are shown.
The spectra have been computed with a statistical approach setting a confidence threshold of 95\% (meaning that the beam displacement spectra will be below the shown curve for the 95\% of the time). 
The control developed for the BPC fully fulfills the RMS requirements, computed in section \ref{sec:accreq}. As it is shown in Figure~\ref{fig:OLCL}, the closed loop RMS (the dashed blue curve) is below the requirement (the dashed black curve) for all four degrees of freedom.
Moreover, the fact that the open and closed loop curves start to overlap above 1~Hz means that the re-introduction of control noise can be considered negligible in the detection bandwidth (above 10~Hz) as it will be described in the following.

\subsection{Beam jitter monitoring: open loop signals}

\begin{figure}[htp]
    \centering
    \includegraphics[width=0.5\textwidth,clip]{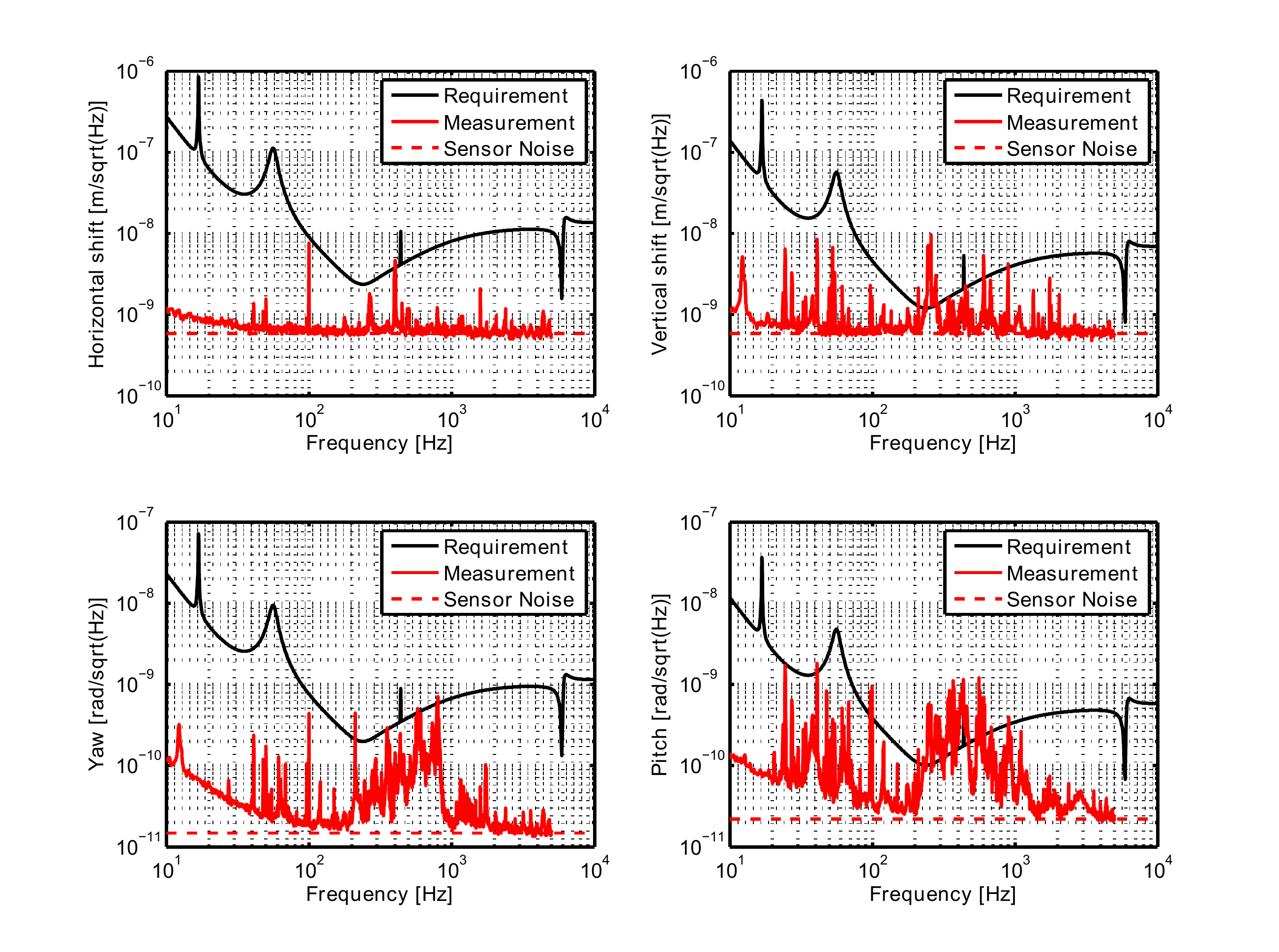}
    \caption{Beam jitter measurement and projection over the detection bandwidth.}
    \label{fig:perfodetband}
\end{figure}

The open loop signals are used to monitor the effective displacement of the beam at the input of the IMC above 10~Hz (see red curves in Figure~\ref{fig:perfodetband}) and have been compared with the Advanced Virgo requirements evaluated in section~\ref{sec:highfrequreq} (black curves).
The sensing is limited by the quadrant sensor noise, dashed red lines in Figure~\ref{fig:perfodetband}, and it is compliant with the Advanced Virgo requirements for all the four d.o.f.. \\
It is worth noting that the displacement measurements performed in the laboratory (red curves in  Figure~\ref{fig:perfodetband})  show structures related to the optical mount resonances excited by the environmental noise of the laboratory, mainly acoustic noise,  which is expected to be lower for Advanced Virgo due to the implementation of acoustic enclosures surrounding the optical bench. 
\section{Conclusion}

In this paper a simple and very effective control system to monitor and suppress the beam jitter noise at the input of an optical system has been described showing the theoretical principle and an experimental demonstration for the application of large scale gravitational wave interferometers.
The control system has shown unprecedented performance in terms of control accuracy and sensing noise. The BPC system has achieved a control accuracy  of $\sim 10^{-8}$~rad for the tilt and $\sim 10^{-7}$~m for the shift and a sensing noise of less than 1~$n$rad resulting compliant with the Advance Virgo gravitational wave interferometer requirements.

\section{Aknowledgements}
The authors would like to thank Flavio Nocera and the EGO electronics group for their support in the quadrant photodiode design, construction and debugging.
The authors would also like to thank Alberto Gennai and Diego Passuelo from INFN-Pisa for having put at our disposal all the control electronics required in this experiment.
We would also like to thank Alain Masserot, Benoit Mours and the LAPP group from Annecy for the Data acquisition system installed in our laboratory which was used in this experiment.
Finally, special thanks go to Fr\'{e}d\'{e}ric Cl\'{e}va from CNRS for his contribution on the initial optical setup and to Richard Day and to Antonino Chiummo from EGO.

\end{document}